\documentclass[%
 aip,
 amsmath,amssymb,
 reprint,%
]{revtex4-1}

\usepackage{graphicx}% Include figure files
\usepackage{dcolumn}% Align table columns on decimal point
\usepackage{bm}% bold math
\usepackage[utf8]{inputenc}
\usepackage[T1]{fontenc}
\usepackage{mathptmx}
\usepackage{textgreek}
\usepackage{color}
\usepackage{subcaption}

\begin{document}

\preprint{AIP/123-QED}

\preprint{AIP/123-QED}
\title[Manuscript Title]
  {Self-starting Harmonic Comb Emission in THz Quantum Cascade Lasers}
\author{Andres Forrer}
\email{aforrer@phys.ethz.ch}
\affiliation{Institute for Quantum Electronics,   Department of Physics,  ETH Z\"urich, 8093 Zürich, Switzerland}
\author{Yongrui Wang}
\affiliation{Department of Physics and Astronomy,
Texas A\&M University, College Station, TX, 77843 USA }
%\author{Martin Franckié}
\author{Mattias Beck}
\affiliation{Institute for Quantum Electronics,   Department of Physics,  ETH Z\"urich, 8093 Zürich, Switzerland}
\author {Alexey Belyanin}
\affiliation{Department of Physics and Astronomy,
Texas  A\&M University, College Station, TX, 77843 USA }
\author{Jérôme Faist}
\author{Giacomo Scalari}
\email{scalari@phys.ethz.ch}

\affiliation{Institute for Quantum Electronics,   Department of Physics,  ETH Z\"urich, 8093 Zürich, Switzerland}

\begin{abstract}
% use sample 2307
Harmonic comb state has proven to be emerging in quantum cascade lasers and promoted by an interplay between parametric gain and spatial hole burning. We report here on robust, pure, self-starting harmonic mode locking in Copper-based double-metal THz quantum cascade lasers. Different harmonic orders can be excited in the same laser cavity depending on the pumping condition and  stable harmonic combs spanning more than 600 GHz bandwidth at 80 K are reported. Such devices can be RF injected and the free running coherence is assessed  by means of self-mixing technique performed at 50 GHz. A theoretical model based on Maxwell-Bloch equations including an asymmetry in the  gain profile is used to interpret the data.
\end{abstract}

\maketitle

%\section*{Introduction}

Quantum Cascade Lasers (QCLs)\cite{faist_quantum_1994, kohler_terahertz_2002} are compact and powerful semiconductor frequency comb sources in the mid-IR \cite{hugi_mid-infrared_2012} and THz regime \cite{burghoff_terahertz_2014,rosch_octave-spanning_2015} exhibiting frequency modulated (FM) output with a linear chirp over the round trip time \cite{singleton_evidence_2018}. They have been investigated in their  properties, stabilization and control in the mid-IR \cite{wang_mode-locked_2009, singleton_evidence_2018} and partially in the THz region \cite{burghoff_terahertz_2014,barbieri_phase-locking_2010, rosch_-chip_2016, mezzapesa_tunable_2019}. Lately, harmonic frequency combs operating on multiples of the fundamental round trip frequency have been reported and investigated in mid-IR Quantum Cascade Lasers (QCLs) \cite{mansuripur_single-mode_2016, piccardo_harmonic_2018, kazakov_self-starting_2017} as well as in actively mode-locked THz QCLs \cite{wang_ultrafast_2020}. Harmonic combs emitted by mid-IR QCLs were observed only after  careful control of the optical feedback \cite{mansuripur_single-mode_2016, piccardo_harmonic_2018, li_dynamics_2015}. Finally, controlled generation of specific harmonic comb states can be achieved by optical seeding \cite{piccardo_widely_2018} or by defect-engineered  mid-IR QC ring lasers \cite{kazakov_shaping_2020}. 

In this Letter, we present  results on robust, self-starting harmonic frequency combs in double-metal Au-Au and  Cu-Cu THz QCLs up to temperatures of 80 K. First, we discuss different THz QCLs which show at multiple bias-points self-starting and pure harmonic comb states and report on their experimentally observed differences with respect  to mid-IR QCL harmonic combs. Successively, the coherence of the spectral modes is examined via a self-mixing technique on the free running electrically detected beatnote. We demonstrate as well injection locking of the harmonic state to an RF synthesizer. The observed harmonic comb state with an asymmetry in the spectral emission is theoretically discussed using a Maxwell-Bloch approach including two lower lasing states. 

%\section*{Harmonic Combs in THz QCLs}

\begin{figure*}
    \includegraphics[width=0.9\textwidth]{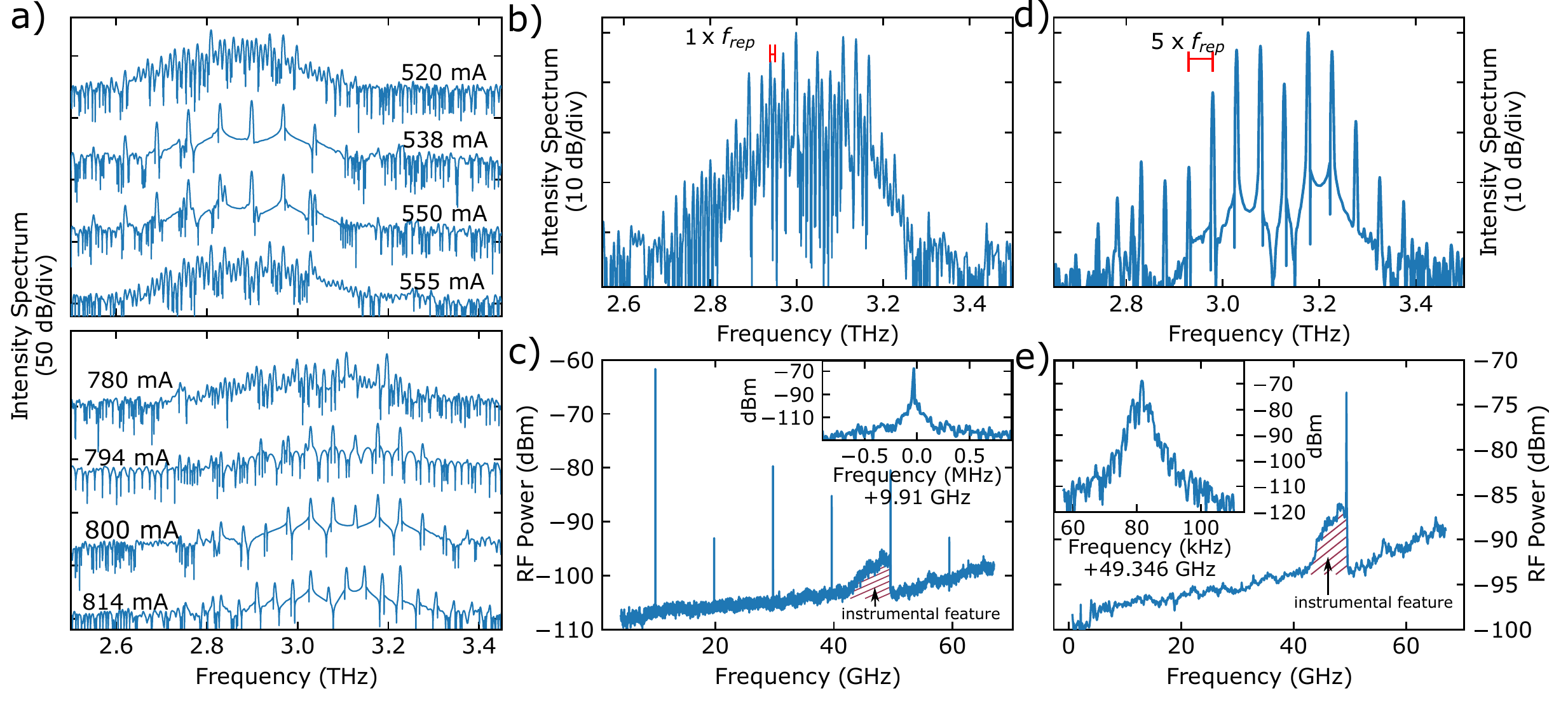}
    \caption{Harmonic comb state in a 4~mm long and 64~$\mu$m wide double-metal Cu-Cu THz QCL at 80 K. a) Transition from a dense state to a harmonic comb state and back for bias current from 520 mA to 555 mA and from a dense state to a harmonic state from 780 mA to 814 mA close to roll-over. Lasing threshold is at 390 mA. b) Intensity spectrum with phase-correction, apodization and zero-padding applied of a fundamental comb state. c) Electrically detected RF spectrum on the bias corresponding to the fundamental comb sate with a narrow beatnote at 9.91 GHz and its harmonics. The increasing noise floor and the transition at 50~GHz is an instrumental feature of the SA used. The inset shows a zoom on the fundamental BN. d) Intensity spectrum of a harmonic comb state with phase-correction, apodization and zero-padding applied. The modes are spaced by 5 $\times$ FSR of the cavity.  e) Electrically detected single beatnote at 49.346~GHz on the laser bias. The absence of other beating signals indicates the purity of the harmonic state. Inset:  Zoom on the harmonic beatnote with a RBW of 300~Hz and a linewidth of 850 Hz with optical feedback from the FTIR. }
    \label{fig:1}
\end{figure*}

\begin{figure}[ht]
    \centering
    \includegraphics[width=0.45\textwidth]{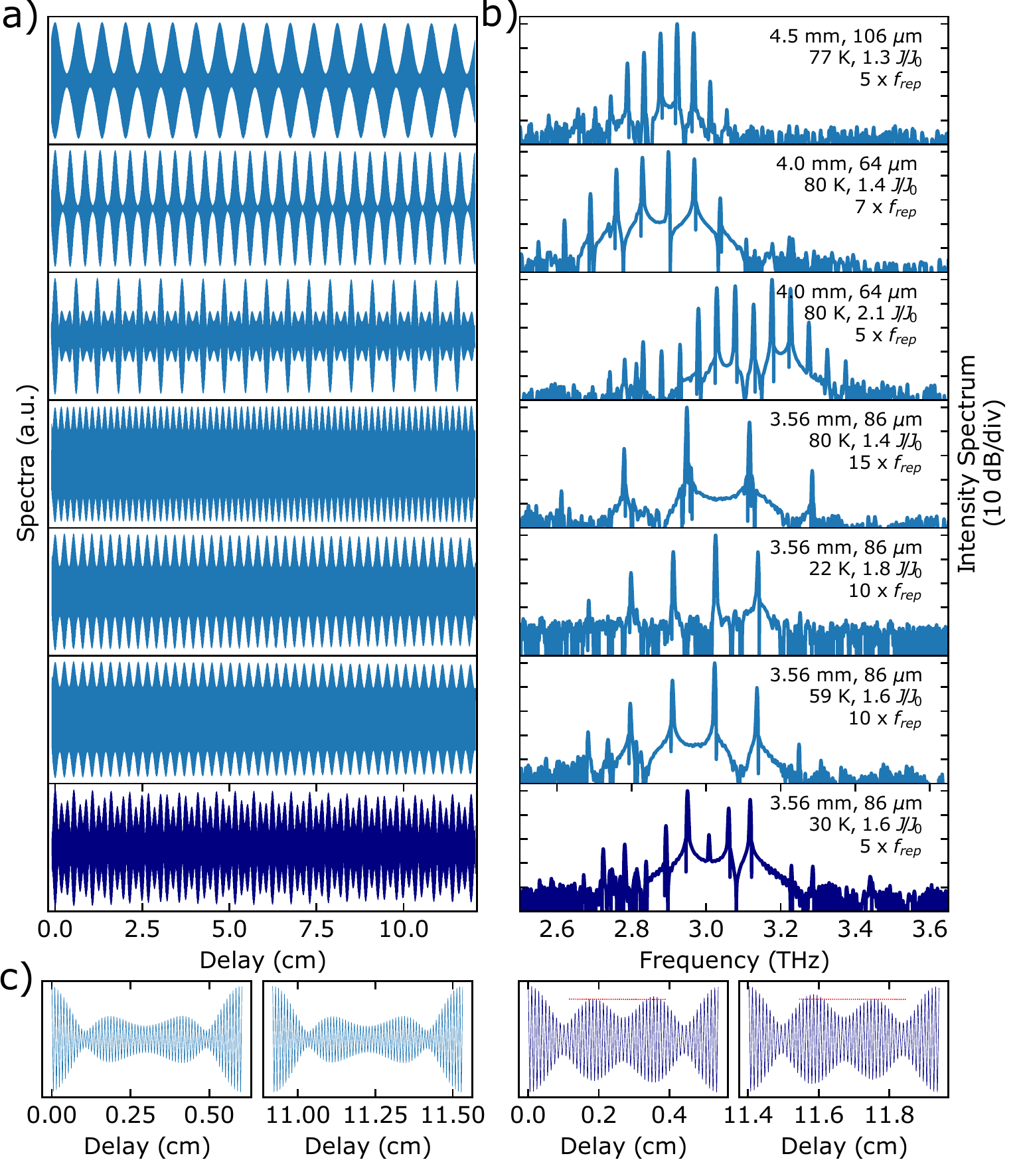}
    \caption{a) shows a series of IFG of harmonic states (light blue) and the harmonic-like state (dark-blue). b) presents the corresponding spectra which all show similarities to harmonic comb states. c) compares the symmetry argument, which is a required property of the harmonic comb (as well for fundamental combs), from the harmonic comb state (light blue) and the harmonic-like state (dark blue), where in the later the symmetry is not preserved.}
    \label{fig:2}
\end{figure}

The investigated THz quantum cascade active region is a homogeneous, four quantum well design based on highly-diagonal transition. The upper-level has two transitions to two lower-levels with dipole moments of $z_{ul_1} ~\sim 3.0$ nm and $z_{ul_2} \sim 2.1$ nm calculated with a Schroedinger-Poisson solver at 30 K. The bandstructure is shown in Fig.~1s of the Supplement. These two transitions lead to an asymmetric gain profile, which significantly changes the theoretical predictions compared to a symmetric gain profile and is discussed in the theoretical part and in detail in the Supplement. The active region itself was previously studied in the fundamental frequency comb regime including injection-locking in Refs. \cite{forrer_photon-driven_2020,forrer_rf_2020}. The active region is embedded in a double metal gold-gold (Au-Au) \cite{forrer_photon-driven_2020} resp. low loss copper-copper (Cu-Cu) \cite{forrer_rf_2020} waveguides featuring lossy setbacks in the top cladding \cite{bachmann_short_2016} for transverse mode control. As was mentioned in Ref. \cite{forrer_photon-driven_2020, forrer_rf_2020} the devices show spectral indications of self-starting harmonic comb states. In the case of Au-Au waveguides lasing starts in a single mode regime and evolves into the harmonic state, finally breaking into the fundamental/dense comb state. This mechanism mainly follows the observations on  mid-IR QCLs \cite{mansuripur_single-mode_2016, kazakov_self-starting_2017, piccardo_harmonic_2018}. In the case of Cu-Cu waveguide we observe a  different behaviour: the harmonic comb does not arise gradually  from a single mode but emerges spontaneously alternating with the fundamental comb or high-phase noise states. The harmonic comb state is observed at different bias-points, even much higher than the laser threshold. An example of a 4.00 mm long and 64 $\mu$m wide device exhibiting self-starting harmonic comb state from a dense state is presented in Fig.~\ref{fig:1}(a) for currents from 520 mA to 555 mA and from 780 mA to 814 mA, operating at 80 K and with a lasing threshold of 390 mA.
The presented spectra are zero-padded, apodized with the Blackman-Harris window, and a phase correcting algorithm \cite{mertz_auxiliary_1967,forman_correction_1966} is applied. %Proper phase correction algorithms like the Mertz- or Froman-algorithm should be applied and help to identify artificial modes since only the positive real part of the complex Fourier Transform (FT) can contain information of the optical intensities in a standard Fourier-transform infrared spectroscopy (FTIR) measurement. Strictly speaking, this is only true for the peak of each mode in the spectrum but not for the noise and therefore for plotting the data in log-scale we take the absolute value of the real part of complex, phase-corrected FT for aesthetic reasons. 
%In a simplified approach the periodicity of the  interferogram (IFG) allows to identify each repeating peak as an "artificial" zero-path-delay (ZPD) in the presence of equally spaced modes and treat the IFG as an effective double-sided IFG (not shown). This allows to simply symmetrically apodize the IFG, perform the complex FT and taking the absolute value but sacrifices all the advantages of phase-correction algorithms.
The symmetry of the interferogram (IFG) arising from equally spaced modes and its interpretation in terms of  coherence is further discussed below and in Fig.~\ref{fig:2}. 
In Fig.~\ref{fig:1}(b) we present the intensity spectrum of a fundamental comb at 755 mA bias current. The electrically detected beatnote at 9.91 GHz is detected over a bias-tee with 67~GHz spectrum analyzer (SA, Rohd\&Schwarz, FSU67) and the RF spectrum is shown in Fig.~\ref{fig:1}(c). The observed linewidth is in the sub-kHz range. The harmonics of the fundamental beatnote resp. the beating of wider spaced modes are visible as well. %It is remarkable that they do not show a decreasing RF amplitude as a function of the frequency.
We observe up to the 6th harmonic (60 GHz) of the fundamental indicating coherence at least up to the 6th mode. 
The observed rise of the noise floor at 50~GHz is an intrinsic effect of the spectrum analyzer (hashed area). The inset shows the fundamental beatnote. Fig. \ref{fig:1} d), measured on the same device,  shows the harmonic comb state skipping four modes at 800~mA driving current. Additionally to the observed THz harmonic spectrum, the intra-cavity mixing of the lasing modes spaced by 5 roundtrips leads to a current modulation  on the bias-line which is as well detected over the bias-tee. The single beatnote at 49.346~GHz is shown in Fig.~\ref{fig:1}(e) corresponding to the expected 5th harmonic of the cavity repetition rate ($49.346/5=9.8692$~GHz).  We would like to highlight here that no other beating across the RF spectrum is observed except the one of the harmonic beatnote, indicating the purity of the harmonic state. The inset of Fig.~\ref{fig:1}(e) shows the strong single beatnote with a resolution bandwidth (RBW) of 300~Hz width a linewidth of $\sim$850~Hz and a $\sim$ 45dB~S/N. The presence of such a strong beatnote at this high frequency (considering also the cables attenuation) could indicate that this harmonic comb is more likely to be caused by an AM instability, see also  Ref. \cite{mansuripur_single-mode_2016}.

Compared to their Au-Au waveguide counterparts, the Cu-Cu waveguide features lower losses (4.4 $cm^{-1}$ resp. 5.7 $cm^{-1}$ at 3~THz for 150 K lattice temperature, 2D Comsol® 5.6 , mirror losses and intersubband absorption excluded.) which lead to increased intracavity fields (roughly 0.2 kV/cm at peak) and therefore increases the effects of parametric gain, discussed in the theory section below. 
This would explain the more frequently observed self-starting harmonic states at different fractions of the threshold current in Cu-Cu devices, which do not only arise from a single-mode instability close to threshold. Multiple of these observed harmonic states appear for different cavity lengths and widths as well as different temperatures. Fig.\ref{fig:2}~a) shows a series of IFGs with the corresponding spectra in Fig. \ref{fig:2}~b). The center of the envelope of the modes, in contrast to mid-IR cases, is found to be in-between two modes.
This asymmetry in the envelope marks a difference in respect to what was observed in mid-IR QCL where a central, more intense mode and symmetric, less intense harmonic modes are frequently observed\cite{piccardo_harmonic_2018}. 

As mentioned above, the coherence of any comb or harmonic state can be verified to a certain degree by comparing the symmetry and periodicity of the measured IFG. The equidistant spacing of the modes in a fundamental or harmonic comb leads to a periodicity of the beating signal, represented by the  delay, in the IFG. This means that for a perfect interferometer all modes are in-phase for delays corresponding to a multiple of the mode spacing ($dx = c/f_{rep}$) and the measured IFG should identically repeat itself. For non-equidistant modes the retardation for being in-phase is different for different mode pairs. Therefore, mode pair beatings are getting out-of-phase for increasing delay. By numerically generating IFGs similar to our case but with non-equidistant spacing we find that a linear increase of $\sim 50$ MHz in the mode spacing still leads to a visible asymmetry, see also Supplement, section VI. This 50 MHz is much smaller than the nominal Fourier Transform resolution of 2.5 GHz of our FTIR. In the case of a comb with more modes this symmetry argument is even valid down to the MHz level. The IFGs in Fig.~\ref{fig:2}(a) exhibit such a symmetry and a periodicity except the last one. It is important to notice that the envelope is slowly decreasing due to diffraction losses and beam divergence of any real FTIR measurement and its alignment but the symmetry within each period is conserved.
The last IFG and spectra in dark blue shows an example of a harmonic-like state that is not  pure. This is shown in Fig.~\ref{fig:2}(c) where a zoomed version of the third IFG from top of the harmonic state (light blue) and the  IFG of the harmonic-like state (dark blue) around the ZPD and close the the maximum travel range of the FTIR are presented. The symmetry is preserved for the harmonic comb whereas for the harmonic-like state a clear asymmetry is observed. This IFG symmetry  argument helps to identify non-pure comb states even in the presence of a single visible electrical beatnote or strong injection. Of course the symmetry in the IFG is a required property for a comb (fundamental or harmonic) and can quantify to a certain extent the coherence between the modes. The coherence should be further tested by SWIFT \cite{burghoff_terahertz_2014}, Intermode Beatnote Spectroscopy \cite{hugi_mid-infrared_2012}, Dual Comb \cite{barbieri_coherent_2011} or any suitable coherence  measurement \cite{CappelliNatPhot2019}. Most of these techniques require fast detectors, but a much simpler approach based on the self-mixing, where the QCL itself acts as a fast heterodyne detector \cite{rosch_-chip_2016}, can verify to some extent the coherence of the modes and was first presented in Ref. \cite{Wienold:14}. 

%\section*{Self-Mixing Intermode Beatnote Spectroscopy}

\begin{figure}[ht]
    \centering
    \includegraphics[width=0.4\textwidth]{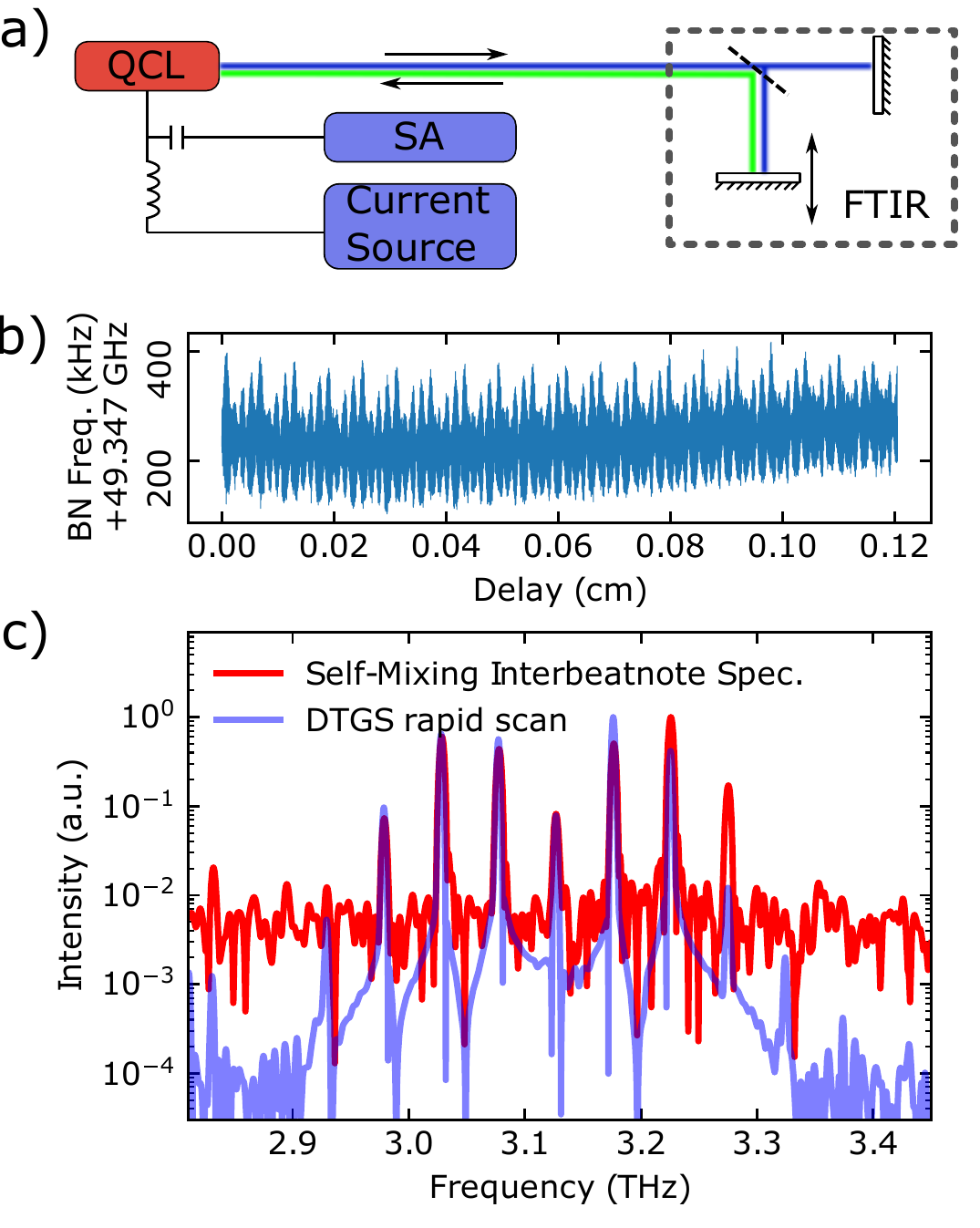}
    \caption{Self-mixing intermode beatnote (SMIB) spectroscopy experiment. (a) sketches the setup where the FTIR provides a slight feedback as a function of time and the induced beatnote frequency shift is recorded by the SA. (b) shows the uncorrected SMIB IFG. The slow drift arises from temperature instabilities. (c) compares the SMIB spectrum with the FTIR spectrum showing the coherence of the modes.}
    \label{fig:3}
\end{figure}

The self-mixing coherence setup is sketched in Fig.~\ref{fig:3}(a) where a 4~mm long and 64~$\mu$m  wide QCL is aligned to a FTIR and the electrical beatnote is detected over the bias-line. It is important to notice that THz devices are intrinsically less sensitive to feedback as compared to mid-IR lasers, due to the high impedance mismatch of the double metal waveguide that provides high facet reflectivity. This allows the use of self-mixing techniques that are not destroying the comb state as observed in the mid-IR \cite{piccardo_harmonic_2018}. The FTIR operates in step-scan where for each step the beatnote intensity and frequency is recorded. From the single mode self-mixing theory of Lang-Kobayashi \cite{lang_external_1980} a weak feedback of the optical mode itself will lead to a slight frequency shift of the optical mode. In the experiment the feedback comes from the FTIR that filters the optical modes as well. In the case of a frequency comb each mode experiences a feedback at its own frequency which will lead to its shift. Since all modes are locked coherently the shift induced to one mode will influence all other modes and their spacing and therefore the beatnote frequency has to be adjusted slightly. This allows one to measure the effect of self-mixing by the frequency change of the beatnote of a laser in a  comb state. A more rigorous mathematical discussion of this concept was first presented in Ref. \cite{Wienold:14}. We already applied this self-mixing intermode beatnote spectroscopy (SMIB) technique  on a single double-metal (Au-Au) THz waveguide emitting simultaneously two unlocked combs spaced by an octave with two independent beatnotes\cite{forrer_coexisting_2018}. We showed that the detected self-mixing signal is sensitive to its generating comb  but not to the other, indicating that indeed only coherent  modes determine the beatnote shift whereas incoherent do not. This approach is applied to the harmonic state presented in Fig.~\ref{fig:1} and the uncorrected SMIB IFG is shown in Fig.~\ref{fig:3}(b) and the spectrum in (c). The induced beatnote change is on the order of $10^{-6}$ of its frequency and the slow drift arise from slight temperature changes. By comparing the SMIB spectra with the DC FTIR spectra we see that all modes are coherently locked and produce the beating signal showing the coherent harmonic comb state. It has to be noted that the SMIB can not fully verify the degree of coherence as it does not provide any information about the relative phases of the modes. 

%\section*{RF Injection Locking}

\begin{figure}[ht]
    \centering
    \includegraphics[width=0.4\textwidth]{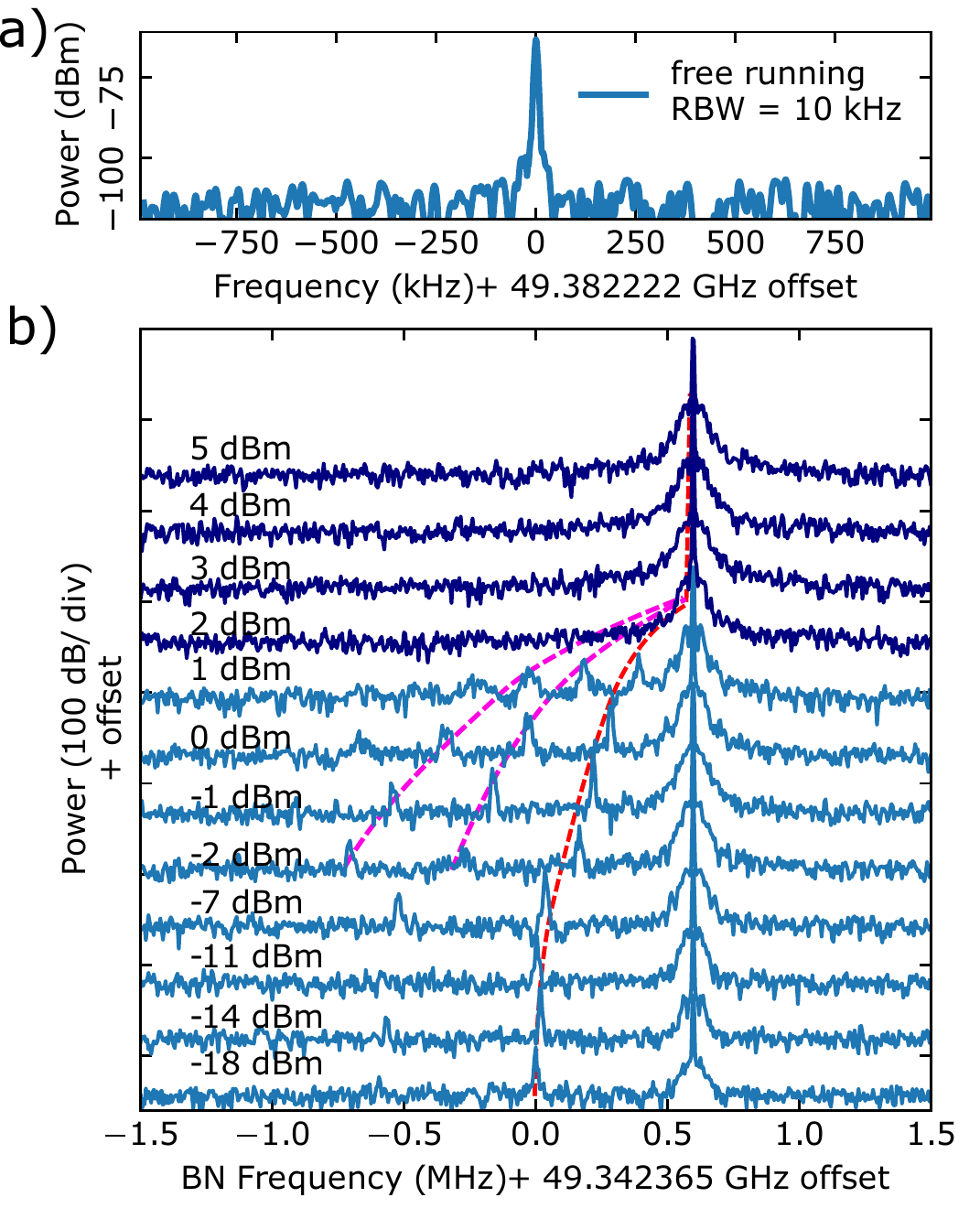}
    \caption{(a): free-running beatnote for the laser operating on the 5th harmonic state. (b): beatnote injection locking as a function of increasing RF power power (from the bottom). Sidebands are also visible and dashed lines are guides to the eye.}
    \label{fig:4}
\end{figure}

%The stabilization of any comb is of interest.
RF injection locking is widely used for comb repetition rate stabilization \cite{RN1028,Hillbrand_NaturePhoton_2019_CoherentInjectionLocking,forrer_photon-driven_2020}. In the following a strong free running narrow beatnote, as shown in Fig.~\ref{fig:4}(a), is injection locked as well. The RF signal is set to a frequency roughly 600 kHz away from the beatnote. While increasing the injection power from -18~dBm at the synthesizer output up to 5~dBm typical pulling, appearance of sidemodes and final locking at 2~dBm is observed as presented in Fig.~\ref{fig:4}(b). The locking range at 2~dBm is roughly 1.2~MHz.

\begin{figure*}[tb]
    \centering
    \begin{subfigure}{0.32\linewidth}
    \includegraphics[width=1.0\linewidth]{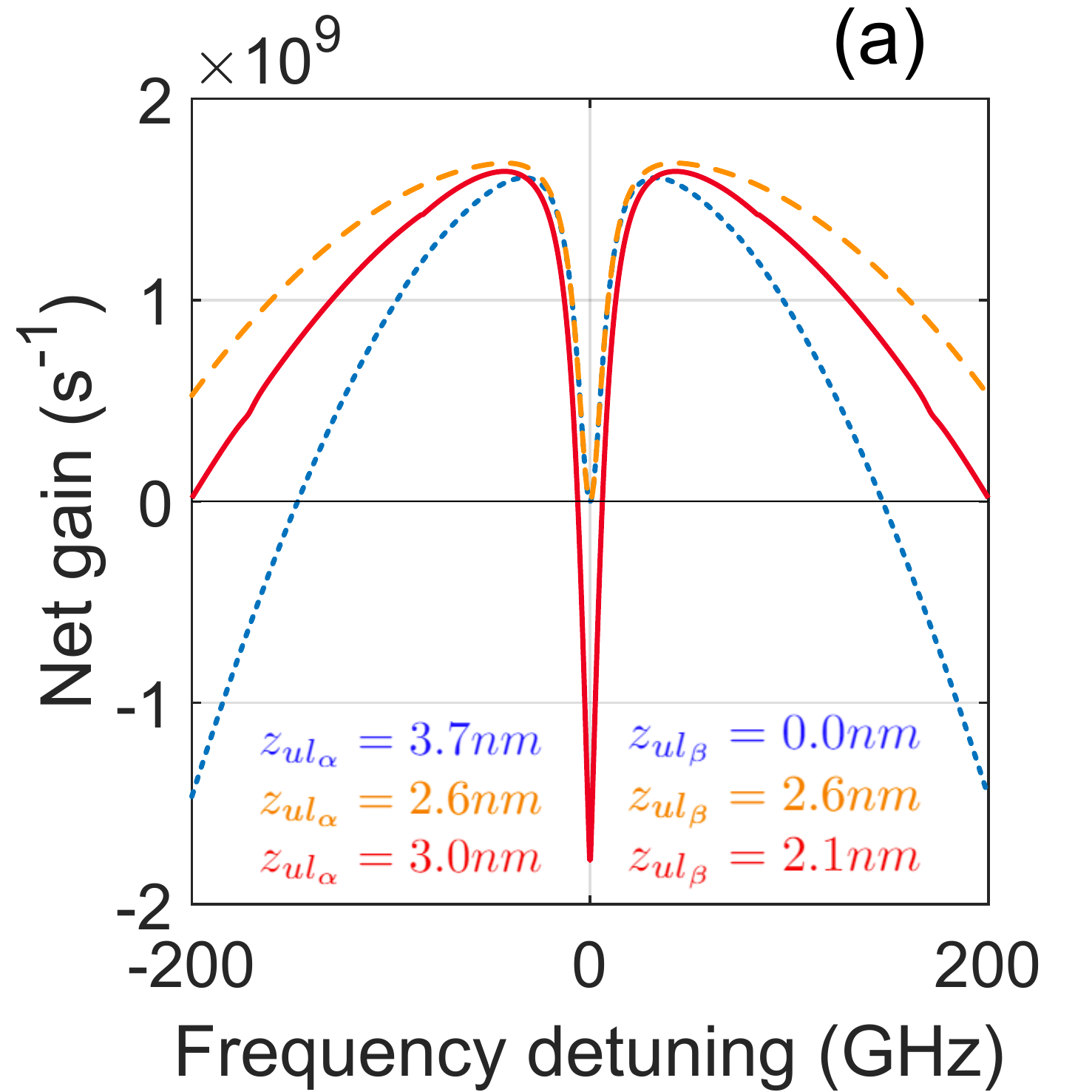}
    \end{subfigure}
    \begin{subfigure}{0.32\linewidth}
    \includegraphics[width=1.0\linewidth]{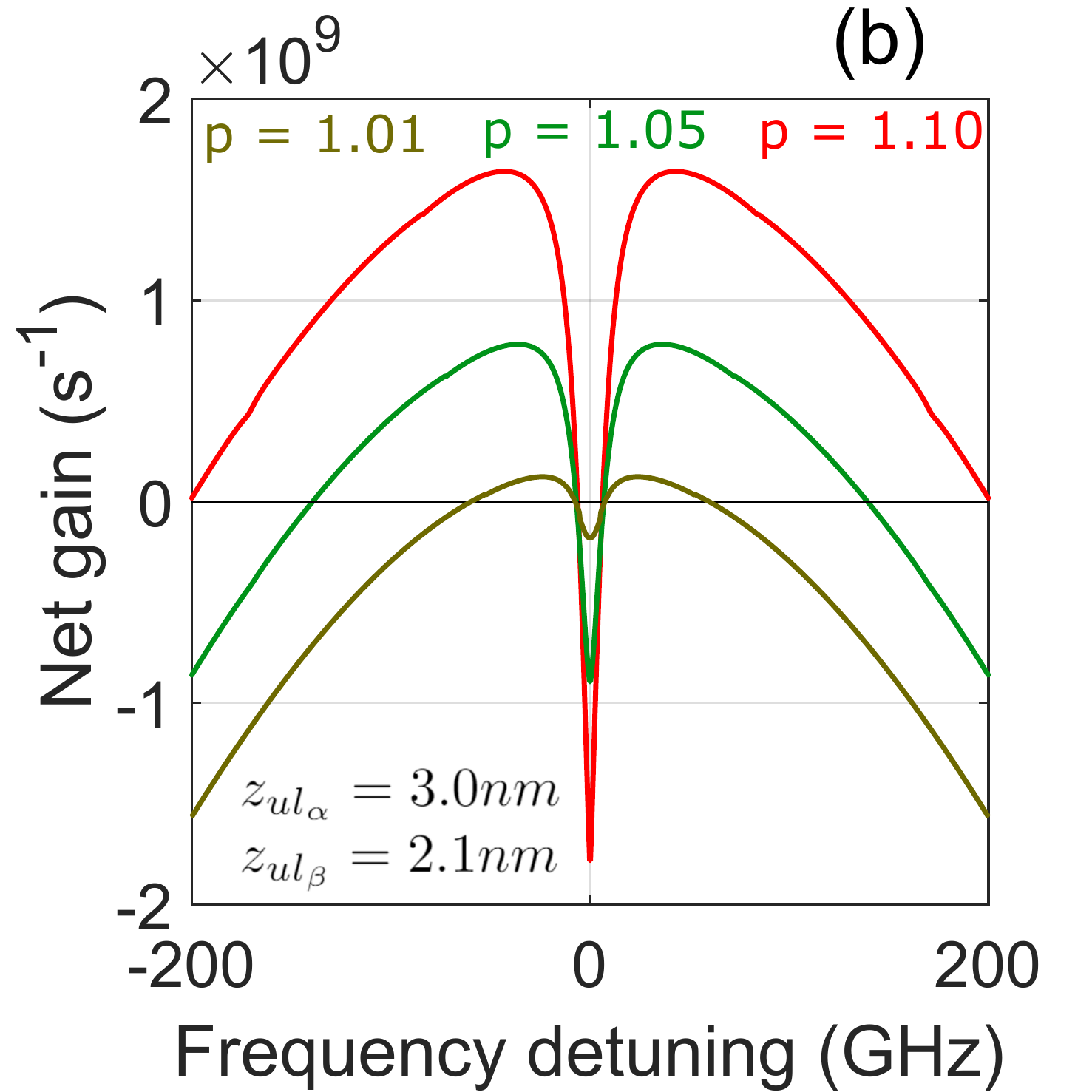}
    \end{subfigure}
    \begin{subfigure}{0.32\linewidth}
    \includegraphics[width=1.0\linewidth]{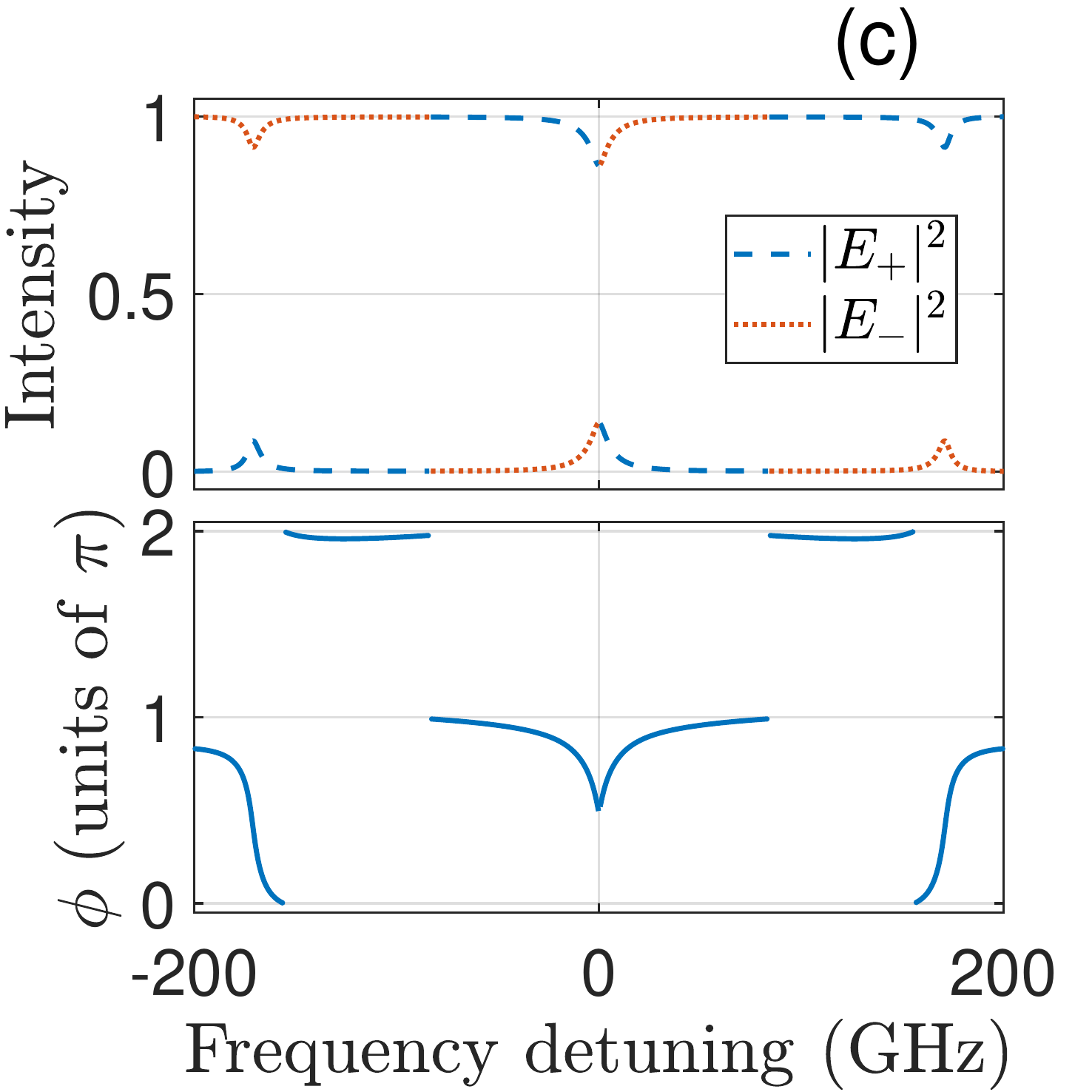}
    \end{subfigure}
    \caption{The instability of weak side modes and their intensities and phases in the presence of a central strong lasing mode as a function of frequency detuning from the central mode for the active region. (a) The net gain at pumping level of $p=1.1$ for single optical transition with dipole moment of 3.7 nm (blue dotted line), two optical transitions with equal dipole moments equal to 2.6 nm (orange dashed line), and two optical transitions with dipole moments equal to 3.0 and 2.1 nm (red line). (b) The net gain for two optical transitions with dipole moments equal to 3.0 and 2.1 nm, at pumping levels of $p$ = 1.01, 1.05, and 1.1, from bottom to top at frequencies larger than 100 GHz. (c) The intensities and phase difference of the weak side modes at pumping level of $p$ = 1.1. The gain relaxation times are $T_1$ = 20 ps and $T_2$ = 0.2 ps, and the population grating diffusion factor is $D = 100 ~cm^2/s$. Horizontal lines in (a) and (b) indicate zero net gain. }
    \label{fig:5}
\end{figure*}

Besides the experimental findings of harmonic combs in THz QCLs and their stabilization and coherence measurement, we develop a theoretical model supporting their formation and appearance. 
We argue that the harmonic comb regime in our lasers can be explained by the interplay of two optical transitions in the active region. This is in fact a common feature of THz QCLs with a diagonal transition design. The presence of two optical transitions with different but comparable dipole moments favors the proliferation of lasing modes separated by several FSRs. To be specific, here we use an active region model which includes one common upper laser state and two lower laser states,
%in agreement with bandstructure calculations for our devices, 
(see Fig.1s of Supplement). Following the approach similar to the one in \cite{wang_harmonic_2020}, we calculate the gain, amplitudes, and phases of weak sidebands generated in the presence of a single strong lasing mode. The details of the calculations and the numerical parameters are described in the Supplement. The formalism is based on the density-matrix equations coupled to the wave equation for the EM field. Since the frequencies of the optical transitions are close to each other, around 11 and 14 meV according to bandstructure calculations, both of them contribute to the optical polarization and laser field. First, we find the field of a strong laser mode in the third-order nonlinear approximation; then we calculate the gain and eigenvalues of weak side modes. At least two side modes have to be included as they are coupled through a strong central mode in the four-wave mixing process.
Figure~\ref{fig:5} presents the net gain, the amplitudes, and the phases of weak side modes for three different cases: (i) only one optical transition with the dipole moment of 3.7 nm;   (ii) two symmetric optical transitions with equal dipole moments equal to 2.6 nm each, and (iii) two asymmetric optical transitions with dipole moments equal to 3.0 nm and 2.1 nm. The dipole matrix elements in the first two cases are chosen so that the values of $d_{ul_\alpha}^2+d_{ul_\beta}^2$ are unchanged from the third case, and hence the total gain coefficients in all cases are similar. From Comsol simulations of waveguide modes, we took the total cavity decay rate corresponding to the field propagation loss of 6.5 cm$^{-1}$ and group velocity dispersion (GVD) of $6.35\times 10^4$ fs$^2$/mm. Comparing the gain spectra in Fig.~\ref{fig:5}(a) one can see that for a single optical transition or two symmetric optical transitions the net gain of weak side modes is positive starting from zero detuning, which indicates that multimode lasing can start from adjacent modes favoring the dense laser spectrum. Only in the case of two asymmetric optical transitions the net gain of side modes is negative at small frequency detunings, preventing lasing on adjacent modes and favoring the harmonic state. 

Figure~\ref{fig:5}(b) shows that the net gain of the weak side modes for two asymmetric optical transitions increases with pumping level at relative large frequency detunings. At the same time, the suppression of gain around zero frequency detuning is stronger at higher pumping levels. This feature does not persist when pumping level is further increased (result not shown), which indicates that harmonic state is more favored at a specific range of pumping levels. 

Fig.~\ref{fig:5}(c) shows that in the case of two asymmetric optical transitions the phase difference between the two side modes is around $\pi$. However, the amplitudes are very different, which indicates that the laser field is not FM, it will have strong amplitude modulation. In contrast, in the case of one optical transition and two symmetric optical transitions, the two weak side modes have similar amplitudes at small frequency detunings ($<$ 30 GHz), and their phase relations indicate FM field. At larger frequency detunings ($>$ 50 GHz), the two side modes have different intensities, which is due to the effect of GVD. See the Supplemental for the details on these two cases.

The results of the linear analysis of the multimode generation for two asymmetric optical transitions are qualitatively consistent with the harmonic lasing state with asymmetric sidebands observed in the experiment. The limitations of weak-sidemode approximation prevent us from making any quantitative predictions. The actual multimode lasing state will be determined by the multiwave mixing of many strong laser laser modes. Its modeling requires fully nonlinear space-time domain simulations  which is beyond the scope of the present paper. Still, the presented analysis allows us to follow analytically how the asymmetric harmonic state emerges from the coherent interplay of two optical transitions, which could be a physical mechanism behind the self-starting harmonic comb emission in THz QCLs. Obviously, more studies are needed before a complete physical picture is revealed. It is fascinating, however, how QCLs continue bringing new surprising features to such a well studied field as fundamental laser dynamics.

\begin{acknowledgments}
We acknowledge the financial support from H2020 European Research Council Consolidator Grant (724344) (CHIC) and from Schweizerischer Nationalfonds zur F\"orderung der Wissenschaftlichen Forschung (200020-165639). Y.W. and A.B. acknowledge the support from NSF grant No. 1807336.  We thank the group of U. Keller for lending the 50 GHz signal generator and thank Martin Franckié for the discussion. 
\end{acknowledgments}

\section*{DATA AVAILABILITY}
The data that support the findings of this study are available from the corresponding author upon reasonable request.



\bibliography{zotero2}

\end{document}